\documentclass[11pt,a4paper]{article}
\usepackage{jcappub}
\usepackage{bm}
\usepackage{latexsym}
\usepackage[latin1]{inputenc}
\usepackage{dcolumn}
\usepackage{amsfonts}
\newcommand{\nab}{\mbox{\boldmath $\nabla$} {}}
\newcommand{\vv}{\mbox{\boldmath $v$} {}}

\newcommand{\zz}{\mbox{\boldmath $z$} {}}
\newcommand{\rr}{\mbox{\boldmath $r$} {}}
\newcommand{\kk}{\mbox{\boldmath $k$} {}}
\newcommand{\VV}{\mbox{\boldmath$V$}}

\title{On the mass of ultra-light bosonic dark matter
from galactic dynamics}

\author{V. Lora$^{a}$, Juan Maga\~na$^{b,1}$, 
\note{Part of the Instituto Avanzado de Cosmolog\'ia (IAC) collaboration http://www.iac.edu.mx/}
Argelia Bernal$^{c,1}$, \\
F. J. S\'anchez-Salcedo$^{b,1}$ and E. K. Grebel$^{a}$}

\affiliation[a]{Astronomisches Rechen-Institut, Zentrum f\"{u}r Astronomie der Universit\"{a}t Heidelberg, \\
             M\"{o}nchhofstr. 12-14, 69120 Heidelberg, Germany}
\affiliation [b]{Instituto de Astronom\'{\i}a,
              Universidad Nacional Aut\'onoma de M\'{e}xico,
              AP 70-264, 04510 Mexico City, Mexico}
\affiliation[c]{Instituto de Ciencias Nucleares,
              Universidad Nacional Aut\'onoma de M\'{e}xico,
              AP 70-543, 04510 Mexico City, Mexico}

\emailAdd{vlora@ari.uni-heidelberg.de}
\emailAdd{jmagana@astroscu.unam.mx}
\emailAdd{argelia.bernal@nucleares.unam.mx}
\emailAdd{jsanchez@astroscu.unam.mx}
\emailAdd{grebel@ari.uni-heidelberg.de}

\abstract{
We consider the hypothesis that galactic dark matter is composed of
ultra-light scalar particles and use internal properties of
dwarf spheroidal galaxies to establish a preferred range for the
mass $m_{\phi}$ of these bosonic particles.
We re-investigate the problem of the longevity of the cold clump
in Ursa Minor and the problem of the rapid orbital decay of the globular
clusters in Fornax and dwarf ellipticals.
Treating the scalar field halo as a rigid background gravitational potential 
and using $N$-body simulations, we have explored how the dissolution
timescale of the cold clump in Ursa Minor depends on $m_{\phi}$.
It is demonstrated that for masses in the range $0.3\times 10^{-22}$ eV 
$<m_{\phi}<1\times 10^{-22}$ eV, scalar field dark halos without
self-interaction would have
cores large enough to explain the longevity of the cold clump
in Ursa Minor and the wide distribution of globular clusters in
Fornax, but small enough to make the mass of the dark halos 
compatible with dynamical limits. 
It is encouraging to see that this interval of $m_{\phi}$ is 
consistent with that needed to 
suppress the overproduction of substructure in galactic halos and
is compatible with the acoustic peaks of cosmic microwave radiation.
On the other hand, for self-interacting scalar fields with
coupling constant $\lambda$, values of
$m_{\phi}^{4}/\lambda\lesssim 0.55\times 10^{3}$ eV$^{4}$ 
are required to account for the properties of the dark halos of
these dwarf spheroidal galaxies.
}


\begin{document}
\maketitle
\flushbottom
\section{Introduction}
The Concordance Cosmological Model, usually referred to as the  
$\Lambda$+ cold dark matter ($\Lambda$CDM) model, has proved 
to be very successful in 
explaining observables across a wide rage of length scales,
from the cosmic microwave background (CMB) anisotropy to the Lyman-$\alpha$
forest.
In this model, nonbaryonic collisionless cold dark matter 
(hereafter CDM) makes up $23\%$ of the total mass content 
of the Universe. 

Observational data on galactic scales seem to disagree with CDM 
predictions. High resolution $N$-body simulations have shown that
the predicted number of subhalos
is an order of 
magnitude larger than what has been observed \citep{klypin,ostriker}.
Another discrepancy arises when comparing the density profiles of
dark halos predicted in simulations with those derived
from observations of dwarf spheroidal (dSph) galaxies 
and Low Surface Brightness galaxies (LSB's); 
$N$-body simulations predict an universal cuspy 
density profile, while observations indicate that a cored halo is
preferred in an important fraction of low-mass galaxies 
\citep{bosch, kleyna03, blok}.

These discrepancies might be overcomed by considering other alternatives.
One intriguing possibility is to consider DM particles 
as spin zero bosons of ultra-light mass $m_{\phi}$,
having such a large Compton wavelength and 
a very large number density that they can be described by a classical 
scalar field $\phi$ (hereafter Scalar Field Dark Matter; SFDM)
\citep{sin, peebles1, peebles2, sahni, hu, matos_2, matos_21, matos_3, arbey1, 
lee2, david, andrew, luis2, hwang,abril}. In this scenario the massive scalar field only 
interacts gravitationally with the rest of the matter and is 
minimally coupled to gravity. Different potentials have been proposed to the 
scalar field, 
e.g.~$V(\Phi)=V_{0}\left[\cosh\left( \xi \Phi \right)-1\right]$ \citep{matos_3},  
$V(\Phi)=m^{2}_{\phi}\Phi^{2}/2$ \citep{turner, magana} and 
$V(\Phi)=m^{2}_{\phi}\Phi^{2}/2 + \lambda \Phi^{4}/4$ \citep{abril}. 
At cosmological scales,  
it is well known that if the quadratic term of $V_{\phi}$ is dominant,  
the SFDM behaves as CDM and thus
linear perturbations of SFDM evolve as those in the standard CDM 
paradigm \citep{turner, magana, abril}.
However, a difference between SFDM and CDM models is 
that a cut-off in its mass power 
spectrum, which can prevent the overproduction of substructures, naturally 
arises in SFDM.  For a quadratic potential, the substructure 
overproduction issue could be overcome if the the boson associated 
to the scalar field is ultra-light with a mass 
of $m_{\phi}\sim 10^{-23}-10^{-22}$ \citep{hu, matos_3}. 
It is important to note that an ultra-light bosonic DM
particle of mass $\sim 1 \times 10^{-22}$ eV is compatible with 
the acoustic peaks of the CMB radiation \cite{ivan}. 

On the other hand, it has been shown that the scalar field can form stable 
structures that could 
account for DM halos \citep{chavanisII}. It is known  \citep{gleiser, tdlee, schunk, seidel,jetzer, guzman, argelia} that ground 
states are stable against spherical and non-spherical perturbations and, furthermore, 
that such configurations are late-time attractors for quite arbitrary 
initial profiles of the scalar field \cite{argelia}. These findings suggest that 
ground states could be naturally formed from initial fluctuations 
of the hypothetical SFDM.  

Interestingly, SFDM halos would have cored mass density profiles.
\citet{arbey1} have analyzed the rotation curves in a sample of LSB galaxies
with SFDM and conclude that ground states can explain fairly well
the observed rotation curves, if the mass of the boson is 
$m_{\phi} \sim 10^{-24} - 10^{-23}$ eV (see also \citep{sin, ji, bernal, jaeweon}).
If the scalar field is self-interacting, its mass $m_{\phi}$ and its
self-coupling parameter $\lambda$ are both constrained in order to
fit the rotation curves of spiral galaxies 
\citep{jaeweon1, arbey2, bohmer, harko11}. 
The dynamics of head-on collisions of ground state halos 
have been also studied \citep{argelia1, paco1}. 

Here we consider the DM halos of dSph galaxies
in the context of SFDM. Our aim is to provide new constraints on SFDM assuming that
it behaves as a massive and complex scalar field.
We will discuss cases with and without self-interaction.
dSph galaxies provide a unique testing ground for the nature of DM.
There is growing evidence that Ursa Minor (UMi), Fornax, Sculptor, Carina, 
Leo I and Leo II possess cored DM 
halos instead of cuspy \cite{kleyna03, goerdt, sanchez1, battaglia,amorisco, 
gilmore}.
It is worthwhile exploring if the SFDM scenario can explain
the existence of cores in dSph galaxies and circunvent some
problems in galactic dynamics in a natural way. 
Here we investigate two of these problems. First of all, we argue
that the interpretation that the stellar clump 
in UMi is a `dynamical fossil' gives constraints on $m_{\phi}$ and $\lambda$.
To do so,
we perform $N$-body simulations of the evolution of a cold stellar clump
embedded in the scalar field DM halo, treating it as a rigid background
potential, but including clump's self-gravity.
Secondly, we argue that, for some combinations of 
$m_{\phi}$ and $\lambda$, the reduction of the 
gravitational dynamical friction in
a SFDM halo could help alleviate the problem of the orbital decay of
globular clusters (GCs) in dSph galaxies and dwarf ellipticals.


The article is organized as follows.
In  \S \ref{sec_SP} we describe the SFDM model and  
briefly review the Schr\"{o}dinger-Poisson system. 
In \S\ref{sec:UMi}, we derive constraints on the values of $m_{\phi}$ 
by studying the puzzling internal dynamics of UMi dSph galaxy.
In \S\ref{Fornax} we discuss the implications of SFDM
in the Fornax dSph galaxy. Finally, in section \S\ref{sec_conclusions} 
we discuss the results and give our conclusions.

\section{The Schr\"{o}dinger-Poisson System} \label{sec_SP}
Since galactic halos are well described as Newtonian systems, 
we will work within the Newtonian limit.
In this limit, the Einstein-Klein-Gordon (EKG) equations 
for a complex scalar field $\Phi$ minimally coupled to gravity and
endowed with a potential $V(\Phi)=m_{\phi}^{2}\Phi^{2}/2+\lambda \Phi^{4}/4$,
can be simplified to the Schr\"odinger-Poisson equations (SP):
\begin{eqnarray}
i\hbar\partial_{t} \psi &=& -\frac{\hbar^2}{2m_{\phi}} \nabla^2 \psi + 
U m_{\phi} \psi    
+ \frac{\lambda}{2m_{\phi}} |\psi|^2\psi \textrm{ , } \label{schroedingerA}\\
\nabla^2 U &=& 4\pi G m^{2}_{\phi} \psi \psi^\ast, \label{poissonA}
\end{eqnarray}
where $m_{\phi}$ is the mass of the boson associated with the scalar field, 
$U$ is the gravitational potential produced by the DM density source 
($\rho=m_{\phi}^{2}|\psi|^2$), $\lambda$ is the self-interacting coupling constant,  
and the field $\psi$ is related to the relativistic field $\Phi$ through 
\begin{equation}
\Phi=e^{-i m_{\phi}c^2t/\hbar}\psi,
\label{Phipsi}
\end{equation}
\cite{ThesisKevin, ThesisArgelia,Friedberg}. 

We are interested in spherical equilibrium solutions to the SP system. 
They are obtained assuming harmonic temporal behavior for the scalar field 
\begin{equation}
\psi(r,t)=e^{-i \gamma t/\hbar}\phi(r).
\label{eq:4.0A}
\end{equation}
In dimensionless variables we have
\begin{eqnarray}
 r&\rightarrow&  \frac{m_{\phi}c}{\hbar}r, \qquad t\rightarrow  
\frac{m_{\phi}c^{2}}{\hbar}t, 
\label{eq:4A}
\end{eqnarray}
\begin{eqnarray}
 \phi\rightarrow \frac{\sqrt{4 \pi G\hbar}}{c^{2}}\phi, \qquad
U\rightarrow \frac{U}{c^{2}}, \qquad \gamma \rightarrow m_{\phi}c^2\gamma,
\label{eq:4.1A}
\end{eqnarray}
and the stationary SP system can be written then as 

\begin{eqnarray}
\partial^{2}_{r}(r\phi)&=& 2 r (U-\gamma) + 2 r \Lambda \phi{^3} \, , \label{S-icA} \\
\partial^{2}_{r}(rU)&=& r\phi^2 \,  \label{P-icA},
\end{eqnarray}
where 
\begin{equation}
\Lambda = \frac{1}{8\pi}\frac{m_p^2}{m_\phi^2}\frac{c}{\hbar^3}\lambda,
\label{eq:lambda}
\end{equation}
and $m_{p}$ is the Planck mass.
In order to guarantee regular solutions,
we require that the boundary conditions at $r=0$ are 
$\partial_rU=0$, $\partial_r\phi=0$ and 
$\phi(0)=\phi_c$, where $\phi_c$ is an arbitrary value. 
On the other hand, we impose that  
\begin{equation}
M=\int^{\infty}_{0} \phi^2 r^2 dr,
\end{equation}
to be a finite number.

The system (\ref{S-icA})-(\ref{P-icA}) can be treated as an   
eigenvalue problem: given a central value $\phi_c$ and a specific value of 
$\Lambda$, there are discrete values $\{\gamma_i\}$ for which the solutions 
$\{\gamma_i,\phi_i(r),U_i(r)\}$ satisfy
the boundary conditions. Solutions can be qualitatively 
differentiated by the number of nodes of the radial function $\phi_i$. 
The solution for which $\phi$ has zero nodes is called the ground state. 
The first excited state is the name for the solution for which $\phi$ 
has one node and so on. 
In this work, we consider only ground states because they are stable under 
gravitational perturbations and thus are more suitable to model 
galactic halos than the unstable excited states \cite{tdlee}. 

\begin{figure}[htp]
\begin{center}
\includegraphics*[width=6.5cm,angle=-90]{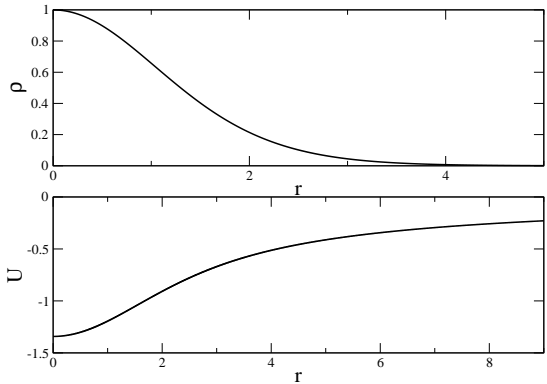}
\caption{Density profile (top) and gravitational
potential (bottom) of the scalar field in a ground state 
with $\phi_{c}=1$ and $\Lambda=0$. 
}
\label{fig:F1}
\end{center}
\end{figure}

Figure \ref{fig:F1} shows the SFDM density profile and the
gravitational potential $U$, as a function of radius $r$, 
for a ground state with $\phi_c=1$ and $\Lambda=0$. 
In order to specify the properties of a certain halo, we will
give the mass of the configuration $M$, and the radius $r_{95}$,
defined as the radius of the sphere that contains $95\%$ of the mass.
In particular, the configuration in figure \ref{fig:F1} has a mass 
$M=2.062$ and $r_{95}=3.799$ in dimensionless units 
(see Eqs.~\ref{eq:4A} and ~\ref{eq:4.1A}).

An interesting characteristic of the SP system is that it obeys the scaling 
symmetry 
\begin{equation}
{\phi,U,\gamma,\Lambda,r}\rightarrow 
{\epsilon^2\phi,\epsilon^2U,\epsilon^2\gamma,\epsilon^2\Lambda,\epsilon^{-1}r}.
\label{scaling}
\end{equation}
This property allows us to find a family of solutions from a 
particular solution. 
Because of this scaling symmetry, the physical quantities of the solutions 
satisfy
\begin{equation}
{M,r_{95}}\rightarrow 
{\epsilon M, \epsilon^{-1}r_{95}}.
\label{scalingMr}
\end{equation}
Therefore, three quantities  
$\phi_{c}$, $m_{\phi}$ and $\Lambda$ or, equivalently $\epsilon$, $m_{\phi}$ 
and $\Lambda$ define a model completely. It is interesting that for 
$\Lambda=0$, $\phi_{c}\sim 10^{-6}$  
($\epsilon \sim 10^{-3}$) and 
$m_{\phi}\sim 10^{-23}$ eV, the ground state would model a halo with 
 $M\sim 10^{10}M_{\odot}$, 
$r_{95}\sim 7$~kpc, and a core radius, $r_{c}$, defined as
the radius where the DM density decays a factor of $2$ its
central value, of $\sim3$~kpc.

While $m_{\phi}$ and $\lambda$ are parameters of the SFDM model that
once fixed they should be universal, 
$\phi_c$ (or equivalently $\epsilon$) is a quantity that may
vary from galaxy to galaxy.  Indeed, it may 
be directly related to the central density of the DM halo. 
A typical value of $\epsilon$ can be inferred from the (dimensionless)
compactness of the ground state defined by 
\begin{equation} 
C=\frac{\epsilon^{2}M}{r_{95}}.
\end{equation}
Assuming that dark halos are Newtonian, it is expected that $C\ll1$. This 
condition is fulfilled if $\epsilon \ll 1$. In fact, studies of 
general relativistic equilibrium solutions, show that configurations
with a very small central value of the scalar field $\Phi_c$ have small 
values of their compactness \cite{PacoLuisSN}. 
Furthermore, it is shown that equilibrium 
configurations with $\Phi_c\lesssim 10^{-6}$  can be treated as Newtonian, that 
is, they can be described within a good approximation by the SP system. 
Because of the relation (\ref{Phipsi}), the condition $\Phi_c\lesssim 10^{-6}$  
implies $\epsilon \lesssim 10^{-3}$.

\section{The halo of UMi dSph: Constraints on SFDM}
\label{sec:UMi}
UMi is a diffuse dSph galaxy located at a distance of $69\pm 4$ kpc \cite{grebel03}    
from the Milky Way center and has a luminosity of
$L_{V}=3\times 10^{5} L_{\odot}$ \cite{grebel03}.
Its stellar population is very old with an age of $10$--$12$ Gyr.
Dynamical studies suggest that UMi is a galaxy dominated by DM,
with a mass-to-light ratio larger than $60 M_{\odot}/L_{\odot}$.
Among the most puzzling observed properties of UMi is that it hosts
a stellar clump, which is believed to be a dynamical fossil that
survived because the underlying DM
gravitational potential is close to harmonic 
\citep{kleyna03,lora}.
This condition is accomplished if the DM halo has a large core.  

We study whether SFDM halos may have cores large enough 
to guarantee the survival of UMi's clump, but small enough to 
ensure that the galaxy is not too massive. Indeed, models with
large cores have large masses. Therefore, we have to impose an upper limit
on $V_{\rm max}$, the maximum of the circular velocity of the halo.
We will see that this provides a stringent constraint on the mass of the boson 
associated with the scalar field and on the self-interaction parameter.

\subsection{The dark halo and the dynamical fossil in UMi}\label{sec_UMi}
As we said before, the most remarkable feature in UMi structure is 
the double off-centered density peak \citep{kleyna98}.
The second peak or clump is located on the north-eastern side of the major 
axis of UMi at a distance of $\sim 0.4$ kpc from UMi's center. 
The velocity distribution of the stars contained in the clump is well fitted
by two Gaussians, one representing the background (velocity
dispersion of $8.8$~km s$^{-1}$) and the other representing the velocity
dispersion of the second peak ($0.5$~km s$^{-1}$). 
The most appealing interpretation is that UMi's clump is a disrupted cluster 
\citep{read} with an orbit in the plane of the sky, which has 
survived in phase-space because the underlying 
gravitational potential is harmonic \citep{kleyna03}, implying
that the dark halo in UMi has a large core.

In order to explore the parameter space of galactic SFDM halos compatible 
with the survival of the kinematically-cold structure in UMi, 
we have studied the evolution of a stellar clump
inside a {\it rigid} SFDM halo in the ground state. 
The dynamics of clump's stars are simulated with $600$ particles in
an $N$-body code. Hence, a star in the clump feels the gravitational
force of a static dark halo potential and the gravity of the remainder of 
the stars in the clump.
In our simulations we ignore the contribution of the stellar background 
component to the gravitational potential because its mass
interior to $\sim 0.4$~kpc ($9\times10^{5}M_{\odot}$) only 
accounts about 9 percent (or less) the DM mass in UMi.

We get the gravitational potential $U$ by solving the SP equations.
In order to specify a SFDM halo we need to fix $\epsilon$ and $m_{\phi}$.
\citeauthor{strigari07} \citep{strigari07,strigari08} 
considered dark halos
compatible with the observed stellar kinematics of the
classical dSph galaxies, including UMi. They found that, for realistic
density profiles, the mass interior
to $300$ pc is $\sim 10^{7}M_{\odot}$ for all dSph galaxies in
the Milky Way halo. Therefore, we fix the $\epsilon$-value
in each SFDM halo by imposing 
that the mass within $r=0.39$ kpc is $1.5\times10^{7}M_{\odot}$
(see also \citep{wilkinson04,penarrubia08,walker09}),
and explore the evolution of the clump for halos with
different $m_{\phi}$.

Initially, the clump has a density profile 
$\rho = \rho_{0}$ exp$(-r^{2}/2r_{0}^{2})$ (with $r_{0}\simeq35$~pc 
\citep{palma}) and 
was dropped at a galactocentric
distance of $0.39$~kpc on a circular orbit in the $(x,y)$ plane
(see also \citep{sanchez2}).
We explored other eccentric orbits but the survival timescale 
is very insensitive to the eccentricity. 
If the $V$-band $M/L$ of UMi stellar population is supposed to be 
$\Upsilon_{\star}=5.8$, the mass of the clump is
$M_{c}=7.8\times10^{4}M_{\odot}$.  The initial 
one-dimensional velocity dispersion of the clump is
$1$~km s$^{-1}$ \citep{sanchez2}. 

The $N$-body simulations were performed with the code described in \citep{lora} and \citep{sanchez2}. This $N$-body code
has the ability to link an individual time step to each particle in the simulation. Only for 
the particle that has the minimum associated time, the equations of motion are integrated. This 
\textquotedblleft multi-step\textquotedblright method reduces the typical CPU times of direct particle-particle
integrations. In our simulations all the particles have the same mass and a softening length of $1$~pc which is 
approximately $1/10$th of the typical separation among clump's particles. 
The convergence of the results was
tested by comparing runs with different softening radii and number of particles.

\subsection{Scalar field without self-interaction ($\Lambda=0$)}\label{resultados_1}

\subsubsection{Upper limit on $m_{\phi}$}
\label{sec:upperlimit}

\begin{figure}
\centerline{\epsfysize=5.5cm \epsfbox{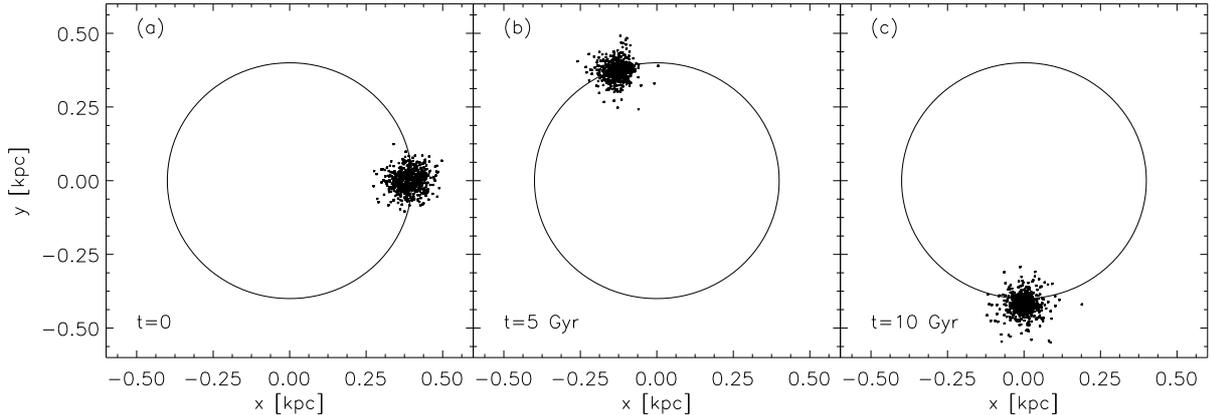}}
\caption{Snapshots of the clump in UMi galaxy, at $t=0$, $5$, and $10$ Gyr.
We set the clump on a circular orbit in the $(x,y)$-plane at a 
distance of $r=0.39$~kpc from UMi's center.
The mass of the boson is $m_{\phi}=10^{-23}$~eV and $\Lambda=0$.
The total mass of the galaxy is $M=9.7\times10^{9}M_{\odot}$.}
\label{fig:F2}
\end{figure}

Figure \ref{fig:F2} shows snapshots of the clump  
at $t=0$, $5$ and $10$~Gyr for $m_{\phi}=10^{-23}$~eV.
The orbit of the clump lies in the $(x,y)$-plane.
We see that the clump remains intact during one Hubble 
time. 
This is a consequence that the SFDM halo has a core large
enough to guarantee the survival of the clump.
For a mass of the boson of $10^{-23}$~eV, the core radius is $2.4$~kpc
and the total halo mass of $9.7\times 10^{9}M_{\odot}$, which is 
too massive (see \S \ref{sec:lowerlimit} for a discussion). 

\begin{figure}
\centerline{\epsfysize=9cm \epsfbox{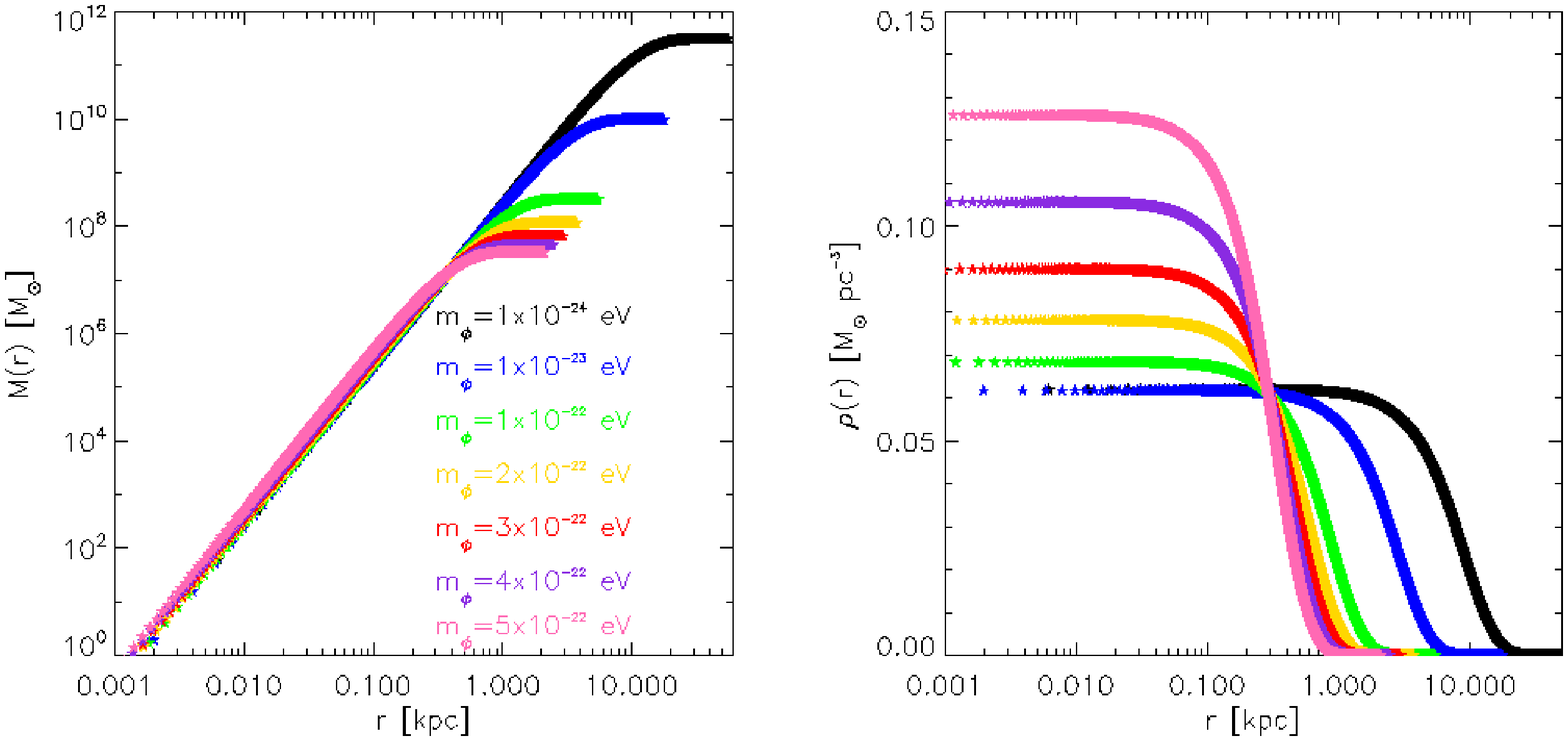}}
\caption{Mass interior to any given radius and 
density profiles for SFDM halos for models with $\Lambda=0$ and
different $m_{\phi}$. }
\label{fig:F3}
\end{figure}

\begin{figure}
\centerline{\epsfysize=5.5cm \epsfbox{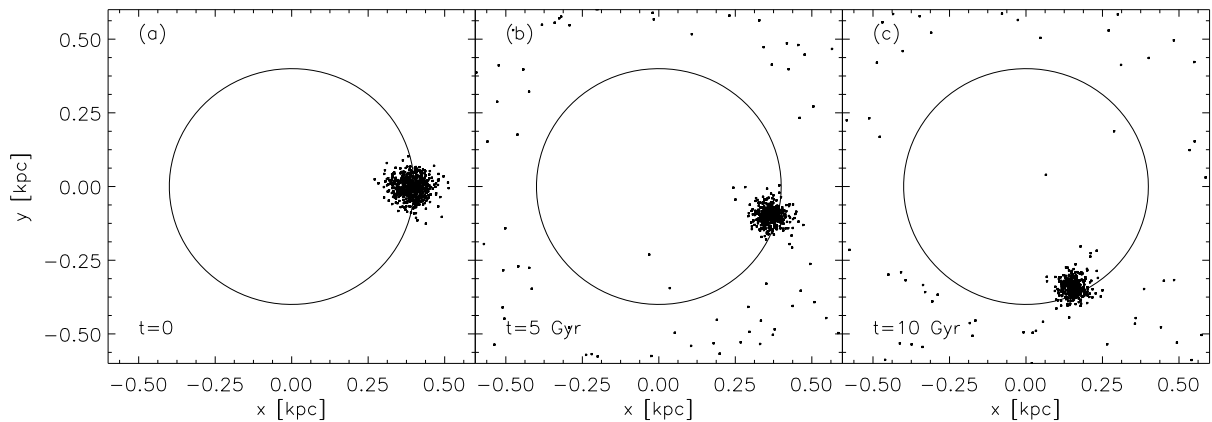}}
\caption{Same as figure \ref{fig:F2} but for a boson mass 
of $m_{\phi}=10^{-22}$~eV and $\Lambda=0$. The UMi's 
mass is of $M=3.1\times10^{8}M_{\odot}$.}
\label{fig:F4}
\end{figure}

\begin{figure}
\centerline{\epsfysize=5.5cm \epsfbox{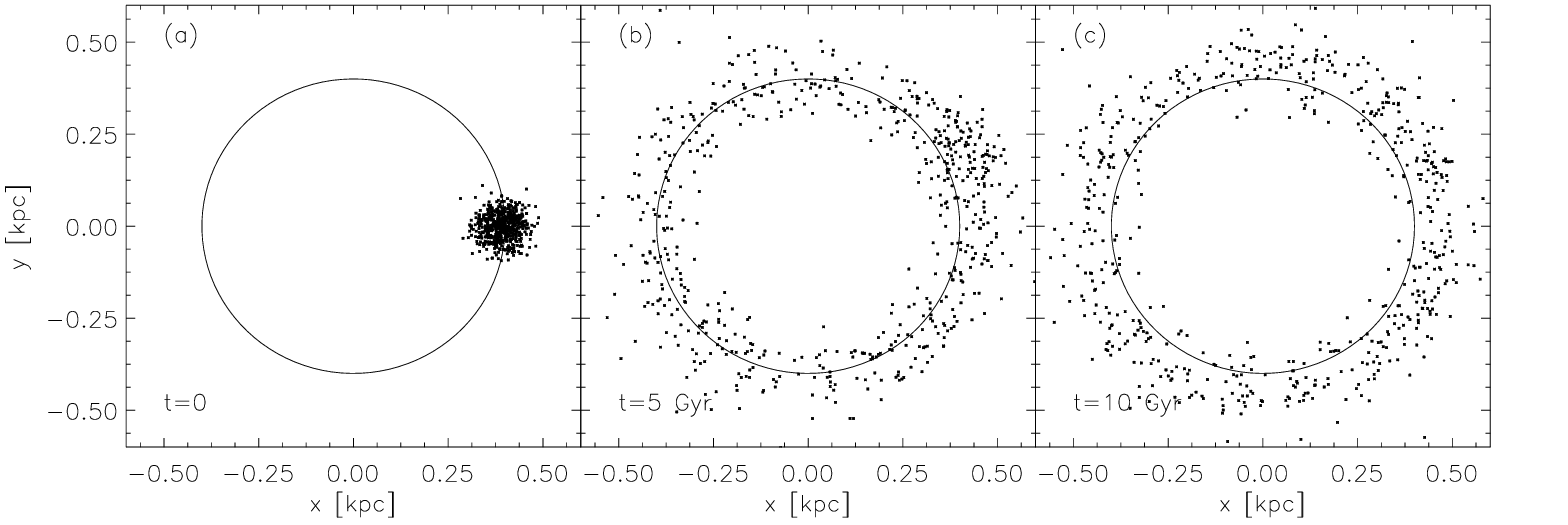}}
\caption{Same as figure \ref{fig:F2} but  
$m_{\phi}=5\times10^{-22}$~eV and $\Lambda=0$. In this case, the total halo
mass is $M=0.33\times10^{8}M_{\odot}$.}
\label{fig:F5}
\end{figure}

As the shape of the underlying gravitational potential depend on
$m_{\phi}$, the longevity of the clump should depend on $m_{\phi}$.
Since both the size of the core and the total mass increase 
when $m_{\phi}$ decreases
(see figure \ref{fig:F3}), we consider now the evolution
of the clump in models with $m_{\phi}$ values larger than $10^{-23}$ eV. 
Figure \ref{fig:F4} shows three snapshots of the clump 
($t=0$, $5$, and $10$~Gyr) in a halo with $m_{\phi}=10^{-22}$~eV, which
has a core radius of $\sim 0.7$~kpc. 
The clump does not dissolve in one Hubble time.
However, when the mass of the boson is $5\times10^{-22}$~eV, 
the core radius of the SFDM halo is quite small, of $0.28$ kpc, 
and, consequently, the clump loses its identity due to tidal effects
(see figure \ref{fig:F5}). 

\begin{figure}
\centerline{\epsfysize=8.5cm \epsfbox{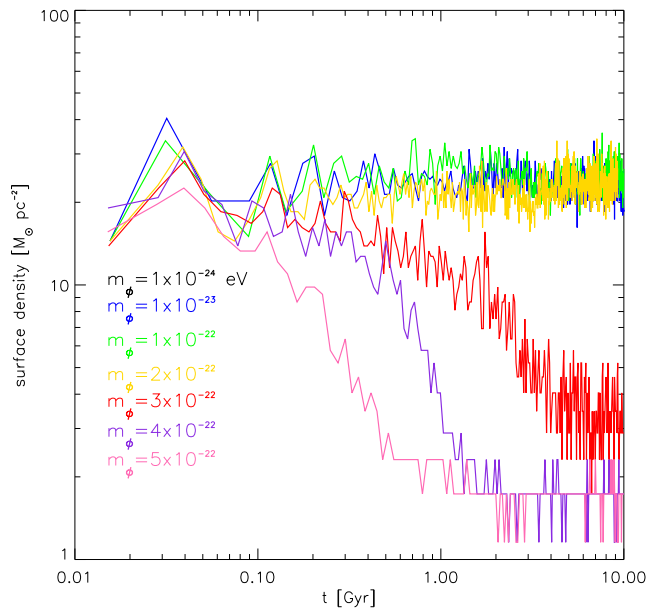}}
\caption{Temporal evolution of $\Pi$ for models with $\Lambda=0$ and
different $m_{\phi}$ values.}
\label{fig:F6}
\end{figure}

To quantify the destruction of the clump in our simulations, 
we calculated a map of the projected surface density of mass 
in the $(x,y)$-plane
at any given time $t$ in the simulation. We sample this two-dimensional map searching for the $10\times10$~pc size
parcel that contains the highest mass, $\Pi(t)$. This parcel is centered at 
the remnant of the clump.
Figure \ref{fig:F6} shows the evolution of $\Pi$ with time for models
with different $m_{\phi}$
and Table \ref{tab:resultados1} summarizes the results of the simulations
with $\Lambda=0$.
We see that in models with $m_{\phi}\sim 3\times 10^{-22}$ eV, the clump is 
diluted within one Hubble time.
In halos with $m_{\phi}>3\times10^{-22}$~eV, the clump is erased
in a too short time.
Therefore, we conclude that the survival of the dynamical fossil sets 
an upper limit to the mass of the boson of $m_{\phi} <3\times10^{-22}$~eV. 

\subsubsection{Lower limit on $m_{\phi}$}
\label{sec:lowerlimit}

In our SFDM models of UMi's halo, there is a positive correlation between
the size of the core $r_{c}$ and the maximum of the circular velocity 
$V_{\rm max}$. While large values of the core are favored to explain
the persistence of the clump and are allowed by the velocity dispersion data,
they require very large values of $V_{\rm max}$ and $M$.
The total mass of the halos are given in Table \ref{tab:resultados1}.
For instance, with $m_{\phi}=10^{-23}$ eV, 
the total dynamical mass is $M=9.7\times10^{9}M_{\odot}$ 
and $V_{\rm max}\simeq 90$ km s$^{-1}$. 
It is extremely unlikely that one of the less luminous classical dSph
galaxies has such a large mass \citep{strigari06}.
Wilkinson et al. \cite{wilkinson04} estimate the enclosed mass of UMi 
within $\sim40'$ ($\sim0.8$ kpc) 
to be of the order of $2\times10^{8}M_{\odot}$ based on projected radial 
velocity dispersion profiles. 
Pe\~narrubia et al. \cite{penarrubia08} suggest that the virial
mass of UMi 
derived from the assumption of a NFW halo model 
is $\sim4.5\times10^{9}M_{\odot}$.
In fact, according to the $\Lambda$CDM simulations by Zentner
\& Bullock \cite{zentner}, 
only $5\%$ of the subhalos in
a Milky Way-sized halo has total masses $> 5\times 10^{9}M_{\odot}$.
We will take this value as the upper limit on UMi mass.
Note that this is a very generous upper limit for $M$; 
fits to the velocity dispersion profile give a mass 
of $2.3_{-1.1}^{+1.6}\times 10^{8}M_{\odot}$ within a tidal radius
of $1.5$ kpc \cite{strigari07}.
If we demand that $M<5\times 10^{9}M_{\odot}$,
we obtain $m_{\phi}>1.7\times 10^{-23}$ eV, whereas for $M<10^{9}M_{\odot}$
we have $m_{\phi}>5\times 10^{-23}$ eV. 
Since there is evidence that at least
three dSph galaxies have a dark halo with a large core 
(Fornax, Sculptor and UMi) and
$\Lambda$CDM simulations have shown that it is unlikely to have 
a population of three subhalos with $M\simeq 5\times 10^{9}M_{\odot}$ 
in a Milky Way-sized halo (see, e.g., \cite{zentner}),
a limit $m_{\phi}>3\times 10^{-23}$ eV seems very reasonable.

Combining the above lower limit with the upper limit derived in \S
\ref{sec:upperlimit}, we find that $m_{\phi}$ should lie between
$0.3\times 10^{-22}$ eV to $3\times 10^{-22}$ eV, being our most
preferred value $m_{\phi}\approx 1\times 10^{-22}$ eV.
For the latter $m_{\phi}$ values, the core radius of UMi dark halo is between
$500$ and $750$ pc.

\begin{table*}
 \centering
 \caption{Destruction times of the clump in UMi dwarf galaxy for 
the $\Lambda=0$ case, for different $m_{\phi}$. The total mass $M$,
and the radius $r_{95}$ and $r_{c}$, are also given for each model.} 
 \medskip
 \begin{tabular}{@{}ccccccc@{}}
 \hline
 $m_{\phi}$ & $\epsilon$  & $M$& r$_{95}$ &  r$_{c}$ & destruction time  &    \\
 ($10^{-22}$eV) & ($10^{-5}$)  &($10^{8}$~M$_{\odot}$) & (kpc) & (kpc) & (Gyr)  &  \\

  \hline
  &  &  &  &  & \\
  $0.01$ &  $111.0$   &$3060$ & $21.8$ & $7.48$ & $>10$ &         \\
  &  &  &  &   \\
  $0.1$ &  $35.1$   &$96.7$ & $6.92$ & $2.37$& $>10$ &              \\
  &  &  &  &   \\
  $1$ &  $11.4$  &$3.14$ & $2.13$ & $0.73$&  $>10$ &                \\
  &  &  &  &   \\
  $2$ &  $8.3$  &$1.15$ & $1.4$ &  $0.50$&  $>10$ &                 \\
  &  &  &  &   \\
  $3$ &  $7.0$  &$0.65$ & $1.15$ & $0.39$ &  $ \sim 5.0$&          \\
  &  &  &  &   \\
  $4$ &  $6.3$  &$0.44$ & $0.96$ & $0.33$ & $\sim 1.5$&           \\
  &  &  &  &   \\
  $5$ & $5.9$   &$0.33$ & $0.82$ & $0.28$& $\sim 0.5$&             \\
  &  &  &  &   \\
 \hline
 \end{tabular}
 \label{tab:resultados1}
 \end{table*}

\subsection{Scalar field with self-interaction ($\Lambda\neq0$)}\label{resultados_2}
So far, we have assumed that boson self-interaction is negligible ($\Lambda=0$).
In order to see how $m_{\phi}$ depends on self-interaction, we 
explored models with the third term of Eq.~(\ref{schroedingerA})
being distinct from zero ($\Lambda\neq 0$).

\subsubsection{Small $\Lambda$-values ($\Lambda \approx 1$)}
\begin{figure}
\centerline{\epsfysize=8.5cm \epsfbox{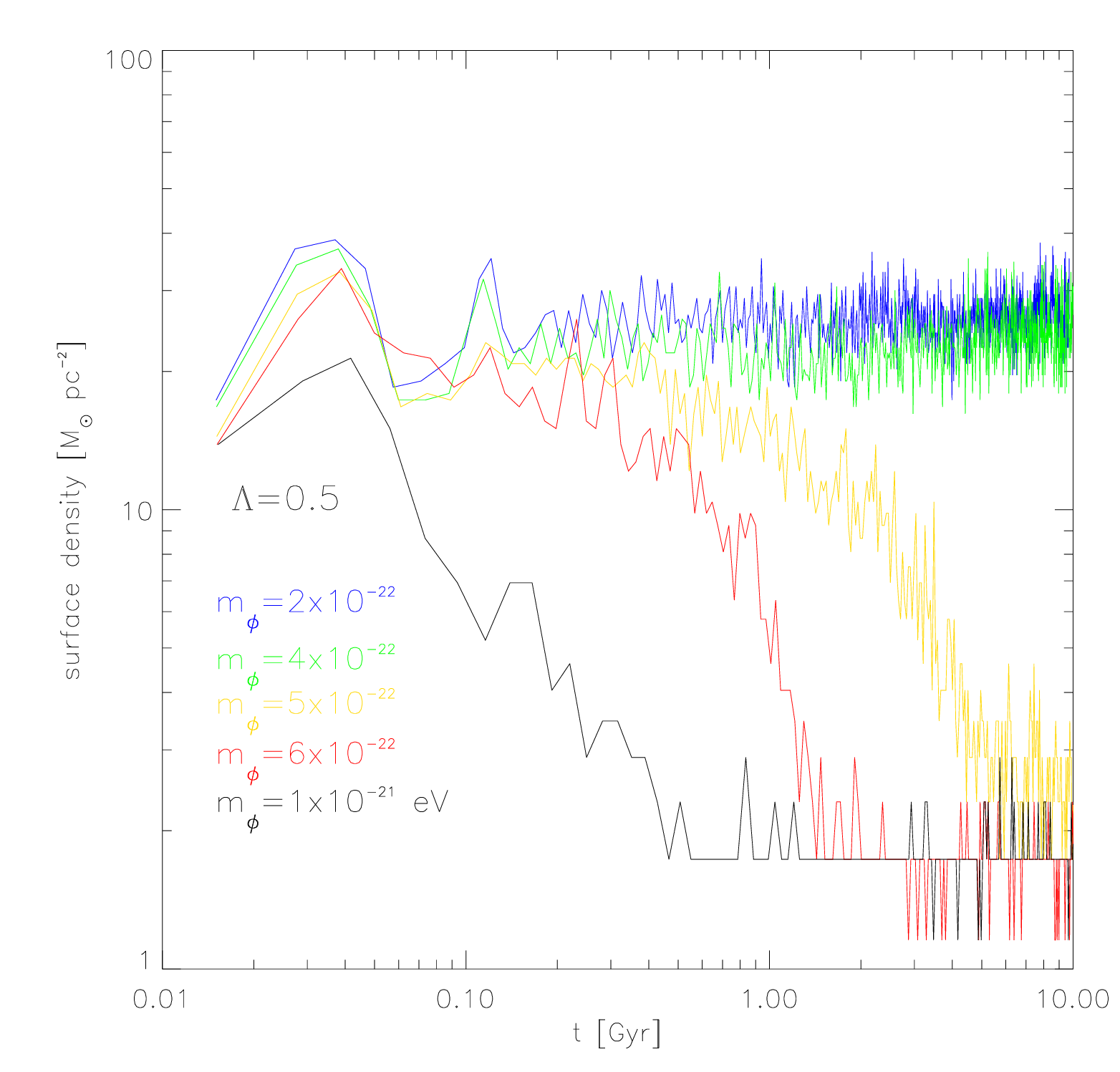}}
\caption{Evolution of $\Pi$ as a function of $t$
for a self-interaction parameter of $\Lambda=1/2$.}
\label{fig:F7}
\end{figure}

\begin{figure}
\centerline{\epsfysize=8.5cm \epsfbox{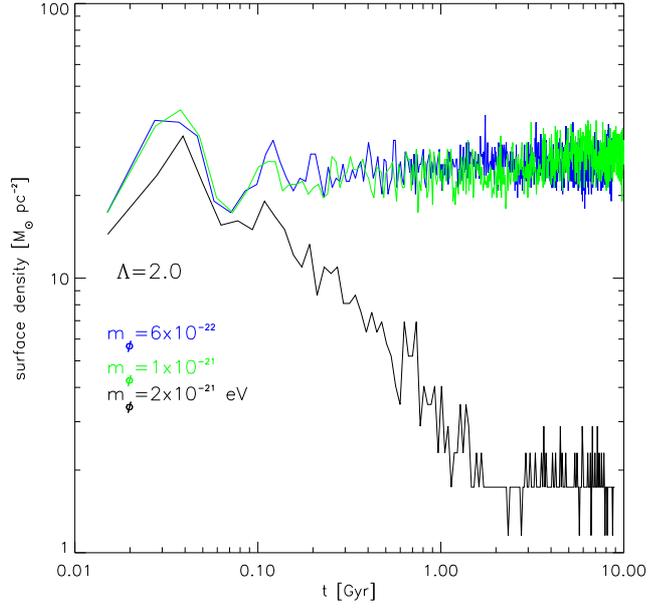}}
\caption{Same as figure~\ref{fig:F7} but for $\Lambda=2.0$.}
\label{fig:F8}
\end{figure}

\begin{figure}
\centerline{\epsfysize=8.5cm \epsfbox{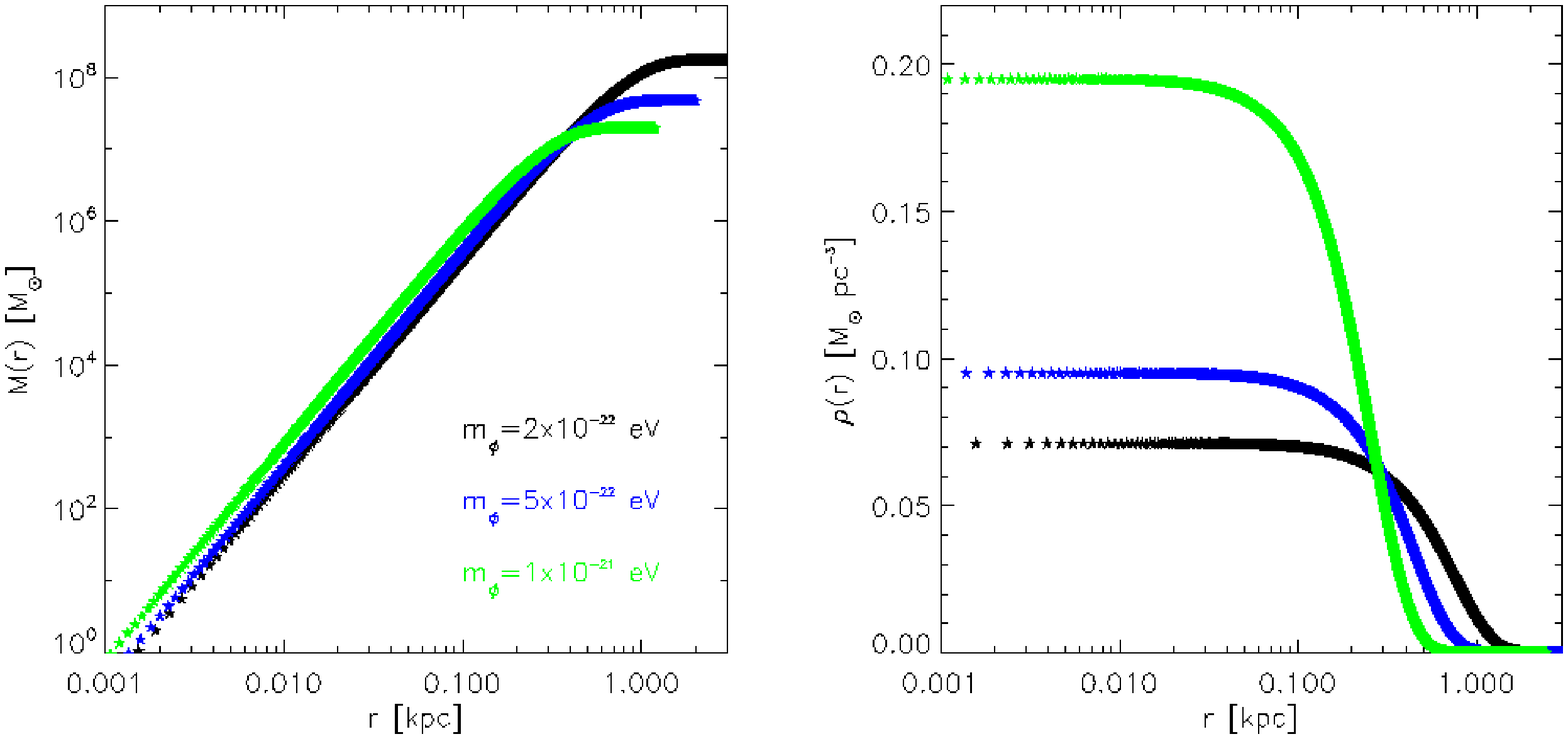}}
\caption{Mass interior to any given radius and 
density profiles for SFDM halos for models with $\Lambda=1/2$ and
different $m_{\phi}$.}
\label{fig:F9}
\end{figure}

\begin{figure}
\centerline{\epsfysize=8.5cm \epsfbox{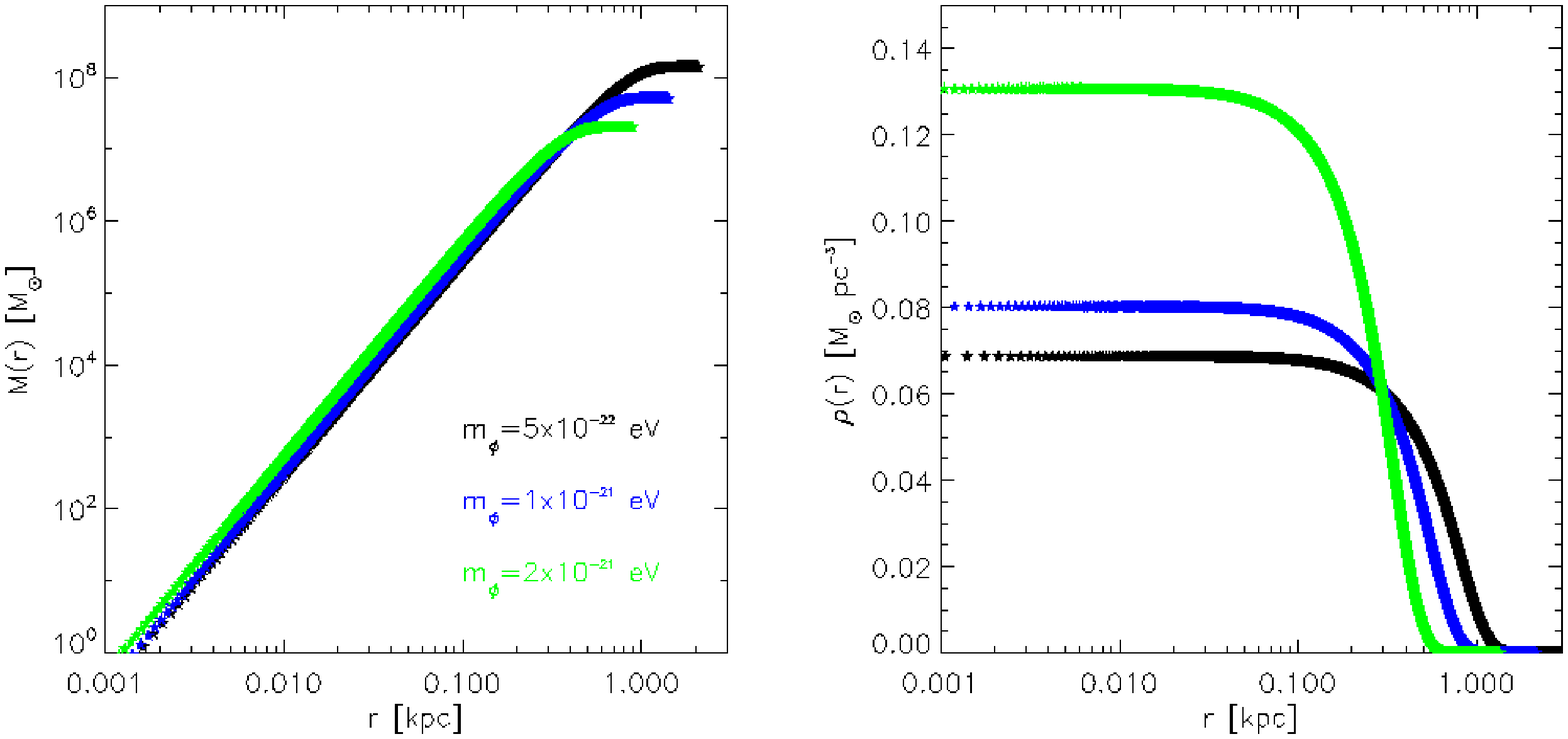}}
\caption{Mass interior to any given radius and 
density profiles for SFDM halos for models with $\Lambda=2$ and
different $m_{\phi}$.}
\label{fig:F10}
\end{figure}

In this section we consider small values 
of $\Lambda$ (in dimensionless units).
The parameters of the models are summarized in Table \ref{tab:resultados2}.
Figures \ref{fig:F7} and \ref{fig:F8} show the evolution of $\Pi(t)$ 
for $\Lambda=1/2$ and for $\Lambda=2$, respectively.
The halo density profiles are given in figures \ref{fig:F9} and \ref{fig:F10}.
With $\Lambda=1/2$, the clump survives even if one increases the mass 
of the boson up to $4\times10^{-22}$~eV 
(here the core radius is $\sim 420$~pc). 
In halo models with $m_{\phi}=1\times10^{-21}$~eV,
the clump is disrupted in $\sim 0.4$ Gyr for $\Lambda=1/2$ but
it survives one Hubble time for $\Lambda=2$. 
As Table \ref{tab:resultados2} shows, for the same $m_{\phi}$,
the core radii of DM halos and their mass increase with $\Lambda$.
With self-interaction, the permitted window for the  mass $m_{\phi}$ 
of the bosonic particles is shifted to larger values. 
We must notice, however, that in order to have $M\gtrsim 10^{8}M_{\odot}$,
as derived by \cite{strigari07}, $m_{\phi}\lesssim 6\times 10^{-22}$ eV
for $\Lambda=2$.  These constraints are compatible
with those found in \cite{rindler11}.

\subsubsection{Large $\Lambda$-values ($\Lambda \gg 1$)}
Here, we analyze the effects on the persistence of the clump of UMi
for large values of $\Lambda$. It was first pointed out in \citep{colpi} 
that the inclusion of self-interaction, even for small 
values of it, can lead to important changes in the resulting scalar field 
configurations. In fact, from the definition of the dimensionless
variable in equation (\ref{eq:lambda}), 
values of $c\lambda/\hbar^3\sim 1$ imply $\Lambda \gg 1$. 
In order to study the system of equations (\ref{S-icA}) and (\ref{P-icA}) 
with $\Lambda \gg 1$, it is useful to rescale the  
variables as follows: $\phi_{*}=\Lambda^{1/2}\phi$ 
and $r_{*}=\Lambda^{-1/2}r$.  
In the limit $\Lambda\to \infty$, 
the Schr\"odinger equation (\ref{S-icA}), to first order
in $\Lambda^{-1}$, has a simple algebraic form
$\phi_{*}=\gamma-U$. Substituting this relation into the 
Poisson equation (\ref{P-icA}) and solving it, we obtain
an exact solution for the scalar density, $\rho_{\phi}=\phi^2$, 
which in terms of the original variables reads as
\begin{equation}
\rho_{\phi} = \frac{\phi_{c}^{2}}{\Lambda} \frac{\sin({\Lambda^{-1/2}r})}{\Lambda^{-1/2}r},
\label{rhosol}
\end{equation}
where $\phi_{c}^{2}$ is an arbitrary constant to be determined. 
The condition $\rho_{\phi}>0$ provides the maximum 
radius $R_{max}$ of the SFDM halo, 
\begin{equation}
 R_{max}= \pi \Lambda^{1/2}.
\label{rmax}
\end{equation}
The mass profile $M(r)$ is given by
\begin{equation}
M(r)=\int_{0}^{r} \phi^2 r'^2 dr'=\phi_{c}^{2}\Lambda^{1/2} \left[\sin \left(\Lambda^{-1/2} r\right) - \Lambda^{-1/2} r \cos\left(\Lambda^{-1/2} r\right)\right], 
\label{M_r}
\end{equation}
and the total halo mass $M$ defined as the mass contained 
within $R_{max}$ reads as \begin{equation}
M=\pi \phi_{c}^{2} \Lambda^{1/2}.
\label{MtotL}
\end{equation}
From dimensionless expressions (\ref{rhosol})-(\ref{MtotL}), 
it is noticed that the free parameters of the halo are $\Lambda$ 
and $\phi_c$. However, when the physical dimensions for those 
expressions are taking into account it turns out
that the free parameters to fit UMi halo are 
$\lambda^{1/2}/m_{\phi}^2$ and $\phi_c$. 
We constrain these parameters through the following steps
\begin{itemize}
\item assume a {\it reasonable} maximum UMi's radius $R_{UMi}$.
Then, use equation (\ref{rmax}) to derive $\lambda^{1/2}/m_{\phi}^2$.

\item given the value of $\lambda^{1/2}/m_{\phi}^2$, $\phi_c$ is 
obtained from equation (\ref{M_r})  by requiring that the mass 
(in units of $m_p^2/m_{\phi}$)
contained within a radius of $0.39$ kpc is $1.5\times 10^7 M_{\odot}$.
 
\item once $\lambda^{1/2}/m_{\phi}^2$ and $\phi_c$ are known, 
the total halo mass and the core radius are calculated
and compared with the dynamical limits discussed in \S \ref{resultados_1}.

\end{itemize}

For instance, if we take UMi's tidal radius of $1.5$ kpc as the maximum 
radius of its
halo, we obtain $m_{\phi}^{4}/\lambda \sim 10^{3}$~eV$^{4}$ 
($\hbar=c=1$) and 
$\phi_{c}^{2}\sim9.6 \times 10^{-9}$. It follows that the total mass is 
$M=3 \times 10^{8}~M_{\odot}$ and the core radius $r_{c}=900$ pc.
This mass is consistent with the 
values derived by \citep{strigari07},  
and the core radius is large enough to 
guarantee the survival of the clump (see tables \ref{tab:resultados1} and 
\ref{tab:resultados2}).
We explored different values for the tidal radius of the halo and 
our results are summarized in table \ref{tab:resultados3}. 
We can see that for a tidal radius of $\sim 1.8$~kpc, 
the clump survives because the core is large enough
($\sim 1$~kpc).  In this case, the SFDM halo becomes a bit more massive.
On the other hand, if we impose the condition  
$M> 1.5\times 10^{8} M_{\odot}$,  
then we obtain $m_{\phi}^{4}/\lambda <1.7 \times 10^3$ eV$^{4}$.

\begin{table*}
 \centering
 \caption{Same as Table \ref{tab:resultados1} but for models with 
small self-interaction parameter.  }
 \medskip
 \begin{tabular}{@{}cccccccc@{}}
 \hline
$\Lambda$ & $m_{\phi}$ & $\epsilon$  & $M$& r$_{95}$ &r$_{c}$&  
destruction time &  \\
 &($10^{-22}$eV) & ($10^{-5}$)  &($10^{7}M_{\odot}$) & (kpc) & (kpc)& (Gyr) &\\

 \hline

  &  &  &  &  & \\
 0.5& $2.$ &  $ 8.13 $  &$17.13$ & $1.57$ & $0.62$& $>10$ &\\

 &  &  &  &  & \\
 0.5& $5.$ &  $5.53$  &$4.66$ & $0.92$ & $0.36$& $\sim 5$ &\\

 &  &  &  &  & \\
 0.5& $10.$ &  $4.68$  &$2.0$ & $0.55$ & $0.22$& $\sim 0.4$ &\\
&  &  &  &  & \\
 \hline

 &  &  &  &  & \\
 2.0 & 2. & $7.9$ & 53.27 & 1.95 & 1.06 & $>10$ & \\ 
  &  &  &  &  & \\
 2.0& $5.$ &  $5.1$  &$13.73 $ & $1.21$ & $0.66$& $>10$  &\\
  &  &  &  &  & \\

 2.0& $10.$ &  $ 3.75$  &$5.05$ & $0.82$ & $0.45$& $>10$ &\\

  &  &  &  &  & \\
 2.0& $20.$ & $3.0$  &$2.02$ & $0.51$ &$0.28$ & $\sim 1.0$ &\\
  &  &  &  &  & \\
 \hline
 \end{tabular}
 \label{tab:resultados2}
 \end{table*}

\section{The halo of Fornax and the orbital decay of GCs} \label{Fornax}
The fact that SFDM halos have cores might solve other apparent problems
in dwarf galaxies. In this section we consider the timing problem
of the orbit decay of GCs in dwarf elliptical galaxies and dSph galaxies. 
In fact, in a cuspy halo, GCs in these galaxies
would have suffered a rapid orbital decay to the center
due to dynamical friction in one Hubble time, forming a nucleated dwarf
galaxy.  For instance, under the assumption that mass follows light or
assuming a NFW profile,
Fornax GCs 3 and 4, which are at distances to the center $<0.6$ kpc,
should have decayed
to the center of Fornax in $\sim 0.5$--$1$ Gyr \cite{goerdt,sanchez1} (see
also Fig. 7 in \cite{ang10}); this clearly represents a timing problem.
Assuming a cuspy NFW halo with the same parameters as those reported 
in \cite{ang10}, 
GCs 1, 2, 3 and 5 can remain in orbit as long as their starting distances
from Fornax center are $\gtrsim 1.6$ kpc, whereas GC 4 needs an initial
distance $\gtrsim 1.2$ kpc.
However, there is no
statistical evidence to suggest that the initial distribution of GCs is so
different to the stellar background distribution.
Lotz et al. \cite{lot01} found that the summed distribution of the
entire GC systems in 51 Virgo and Fornax cluster dwarf
ellipticals closely follows the exponential
profile of the underlying stellar population.
In addition, studies of the radial distribution of GCs in giant
elliptical galaxies show that the distribution of metal-rich GCs
matches the galaxy light distribution \cite{harris,strader}.
Assuming that GCs formed
along with the bulk of the field star population in dwarf galaxies,
the probability that Fornax GCs were formed all
beyond $1.2$ kpc is $\sim (0.03)^{5}\simeq 2.5\times 10^{-8}$ 
(here we have used that the fraction of the stellar mass beyond $1.2$ kpc
for a King model with a core radius of $0.4$ kpc and a concentration 
parameter of $0.72$, as used in \cite{ang10}, is less
than 3 percent).  Therefore, 
it is very unlikely that all the GCs in Fornax were formed at such large 
distances and even they did, there is still a timing problem with GCs 3 and 4.

The dynamical friction timescale depends on the nature of DM particles.
Goodman \cite{goodman} was the first to suggest that the
suppression of dynamical friction in superfluid DM halos could
circumvent some problems in galactic dynamics such as the presence
of rotating bars in dense dark halos (see also \cite{good11,sle11}).
Here we will study if SFDM halos can solve or alleviate the timing problem 
of the GCs in dwarf galaxies.

\begin{figure}
\centerline{\epsfysize=8.5cm \epsfbox{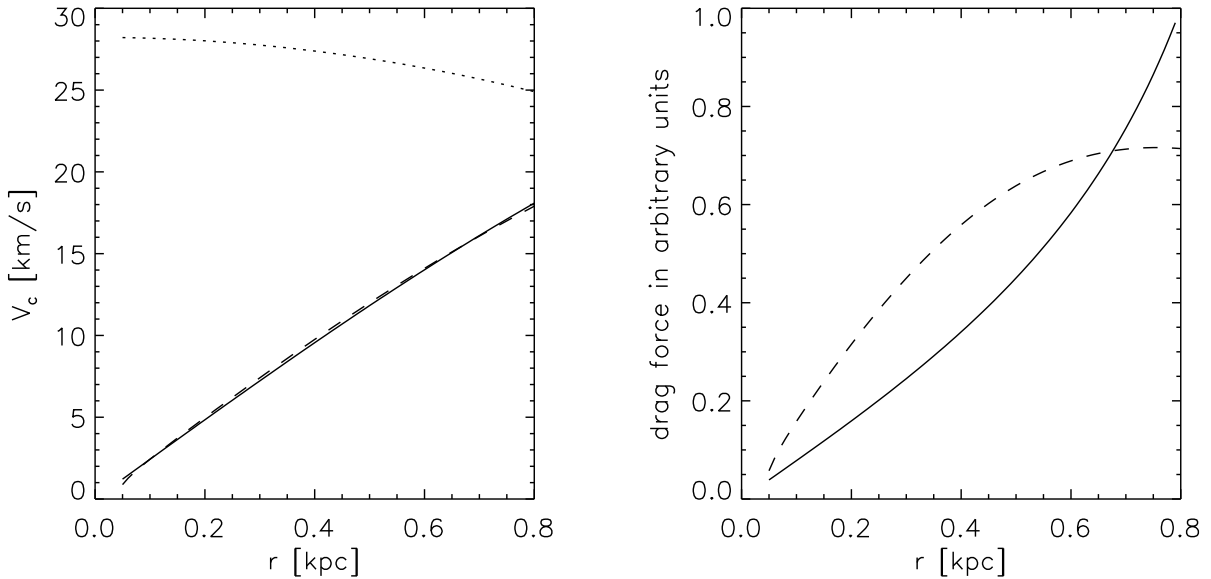}}
\caption{Left panel: Circular velocity (solid  line) and
sound speed (dotted line) as a function of galactocentric distance
for a SFDM model with $m_{\phi}^{4}/\lambda=0.55\times 10^{3}$ eV$^{4}$
and $R_{\rm max}=2.1$ kpc.
For comparison we also plot the circular velocity in the cored 
collisionless DM halo 
employed by S\'anchez-Salcedo et al. \cite{sanchez1} to solve the problem
of orbit surviving. Right panel: Gravitational drag force 
in these models. Solid line stands for the SFDM model and dashed line
for the the cored collisionless halo. }
\label{fig:Fornax1}
\end{figure}

We consider first the so-called Thomas-Fermi regime, that is,
the limit of large scattering parameter ($\Lambda \gg 1$). 
In this limit, the dynamics of the scalar field is 
effectively the same as that for an ideal fluid
with equation of state $P=K\rho^{2}$ and $K=\lambda/m_{\phi}^{4}$
(see \cite{sle11} and the Appendix). 
Therefore, the SFDM dynamical friction force exerted on a gravitational object 
is the same as it is in a gaseous medium with sound speed 
$c_{s}=(2K\rho)^{1/2}$ (see the Appendix).

Ostriker \cite{ostrikereve} calculated the drag force on
a perturbing body of mass $M_{p}$ moving at constant speed $V_{0}$
through an ideal fluid with sound speed $c_{s}$.
She found that for subsonic perturbers,
that is ${\mathcal{M}}<1$ where the Mach number is defined
as ${\mathcal{M}}\equiv V_{0}/c_{s}$, the drag force is
generally smaller in a gaseous medium than in a collisionless medium,
because pressure forces in a gaseous medium create a symmetric
distribution in the vicinity of the perturber.
More specifically, she found that at ${\mathcal{M}}<1$ the dynamical
friction force $F_{DF}$ in a gaseous and homogeneous medium is
\begin{equation}
F_{DF}=\frac{4\pi (G M_{p})^{2}\rho_{0}}{V_{0}^{2}}
\left[\frac{1}{2}\ln\left(\frac{1+{\mathcal{M}}}{1-{\mathcal{M}}}\right)
-{\mathcal{M}}\right],
\label{eq:ostriker_sub}
\end{equation}
where $\rho_{0}$ is the unperturbed fluid density.
Moreover, Kim \& Kim \cite{kimkim} demonstrated that Ostriker's formula
for the subsonic case, which was derived for perturbers in rectilinear 
orbits, is also valid for objects on circular orbits.

We consider now the decay of GCs on
near-circular orbits in a spherical self-gravitating
halo of fluid DM with polytropic index $\gamma=2$, due to dynamical friction. 
In that halo, the circular
speed depends on the galactocentric distance $r$,
$V_{0}(r)=\sqrt{GM(r)/r}$, where $M(r)$ is given in Eq.~(\ref{M_r}).
The motion is
subsonic, i.e.~$V_{0}/c_{s}<1$, at any radius interior to $0.84r_{c}$.
Figure \ref{fig:Fornax1} shows the halo circular velocity, sound speed and 
the drag force (using Eq.~\ref{eq:ostriker_sub}) as a function of the distance
to the center of Fornax for a model with
$R_{\rm max}=2.1$ kpc, which corresponds to $r_{c}=1.25$ kpc and
$m_{\phi}^{4}/\lambda=0.55\times 10^{3}$ eV$^{4}$.
For comparison, we also plot the corresponding
curves for a King halo made of collisionless particles
with a core large
enough to solve the timing problem in Fornax (see \cite{sanchez1} for
details). Within a galactocentric
distance of $0.8$ kpc, the dynamical friction in a collisionless halo with
a King radius of $1.5$ kpc is similar to the drag force in a SFDM
halo with a core radius of $1.25$ kpc and thus, conclude that
models with $m_{\phi}^{4}/\lambda\lesssim 0.55\times 10^{3}$ eV$^{4}$
are compatible with the radial distribution of the population of
GCs in Fornax.

In the noninteracting case, the equation governing the evolution
of the perturbation is no longer a wave equation with constant sound speed
but, instead, the velocity group depends on the wavenumber \citep{chavanis1}.
The derivation of the steady density wake
in a homogeneous medium is outlined in the Appendix.
For a perturber moving at speed $V_{0}$ in rectilinear orbit along 
the $z$-axis, it is convenient to define the 
transonic wavenumber as $k_{0}\equiv m_{\phi}V_{0}/\hbar$ and
the radial wavenumber $k_{R}^{2}=k_{x}^{2}+k_{y}^{2}$.
In the Appendix, we show that Fourier modes with radial wavenumber
$k_{R}$ larger than $k_{0}$ behaves
subsonically, whereas modes with $k_{R}<k_{0}$ are supersonic. 
Consider a GC on circular orbit with radius $R_{\rm orb}=0.8$ kpc 
from the galactic center of Fornax with  
a circular velocity $\simeq 18$ km s$^{-1}$
(see Fig.~\ref{fig:Fornax1}). For $m_{\phi}=(0.1-1)\times 10^{-22}$ eV, the transonic
wavenumber is $k_{0}=0.1-1$ kpc$^{-1}$.
Small $k_{0}$ values imply that the motion is supersonic only
for Fourier modes with large wavelength. Since GCs
are assumed to be on quasicircular orbits, the direction of $\VV_{0}$
changes by $\pi/2$ four times per orbit and thus the forward-wave
propagation effectively restarts as well \cite{ostrikereve,san01,kimkim}. 
This implies that modes
with wavelength much larger than $R_{\rm orb}$ are not relevant.
In fact, because the wake bends behind the body,
the drag force promptly saturates to a steady-state value
within less than the crossing time of modes over the distance
equal to the orbital diameter \cite{kimkim}.
Therefore,
we may state that the perturber is essentially subsonic provided
that $k_{0}^{-1}\gtrsim R_{\rm orb}$.
As dynamical friction is effectively suppressed for subsonic perturbers,
Fornax GCs within a radius of $\sim 1$ kpc are expected to
avoid significant orbital decay if $k_{0}\lesssim 1$ kpc$^{-1}$, which implies
$m_{\phi}<1\times 10^{-22}$ eV.
Thus, $m_{\phi}$ values between $0.3\times 10^{-22}$ eV and
$1\times 10^{-22}$ eV could explain the survival of the dynamical fossil
in UMi and may alleviate the timing problem of the GCs orbiting
dwarf galaxies.

\begin{table*}
 \centering
 \caption{Relevant parameters of the halo for
models with large self-interaction.}
 \medskip
 \begin{tabular}{@{}ccccccccc@{}}
 \hline
$R_{max}$ &$m_{\phi}^4/\lambda$& $\phi_c^2$&$r_{c}$&$M$&\\
(kpc)    &($10^3 eV^4$) &($10^{-9}$) &(kpc)& ($10^{8}$~M$_{\odot}$)&\\ 
 \hline
  &  &  &  &   \\
1.2   &1.7   &6.35   &0.72   &1.6\\
  &  &  &  &   \\
1.3   &1.4   &7.35   &0.78   &2.0\\
  &  &  &  &   \\
1.4   &1.2   &8.4   &0.84   &2.5\\
  &  &  &  &  & \\
1.5   &1.1   &9.6   &0.90    &3.0\\
  &  &  &  &  & \\
1.6   &0.94   &10.8   &0.96   &3.6\\
  &  &  &  &  & \\
1.7   &0.83   &12.1   &1.02   &4.3\\
  &  &  &  &  & \\
1.8   &0.74   &13.5   &1.08   &5.1\\
\hline
\end{tabular}
\label{tab:resultados3}
\end{table*}

\section{Concluding Remarks} \label{sec_conclusions}
We have considered a model where 
ultra-light bosons are the main components of the
dark halos of galaxies.  The main goal of this
work was to constrain the mass of the scalar particles.
We constructed stable equilibrium configurations of SFDM in the
Newtonian limit to model the DM halo in UMi. We studied two 
relevant cases of SFDM halos: with and without self-interaction.

The persistence of cold substructures in UMi places upper limits on $m_{\phi}$.
Using $N$-body simulations, we found that the survival of cold substructures
in UMi is only possible if $m_{\phi} < 3\times10^{-22}$~eV in the
$\Lambda=0$ case. 
On the other hand, by imposing a plausible upper
limit on $M$, we place lower limits on $m_{\phi}$. 
All together, we find that for $\Lambda=0$, $m_{\phi}$ should be in 
the window
\begin{equation}
0.3\times 10^{-22}{\rm eV}< m_{\phi} < 3\times 10^{-22}{\rm eV}.
\label{eq:range}
\end{equation}
Since the timing problem of the orbital decay of the GCs in
Fornax can be alleviated if $m_{\phi}<1\times 10^{-22}$ eV for $\Lambda=0$,
our most favored value is around $(0.3-1)\times 10^{-22}$ eV.

For SFDM models with self-interaction, the upper limit on $m_{\phi}$
increases with $\Lambda$. 
For $\Lambda=2$, halos made up by bosons of mass 
$\lesssim 6\times 10^{-22}$~eV 
could account for the observed internal dynamics of UMi.
In the limit $\Lambda\gg 1$, 
we find that $m_{\phi}^{4}/\lambda\lesssim 0.55\times 10^{3}$ eV$^{4}$
would explain the longevity of UMi's clump and the 
surviving problem of GCs in Fornax.

The window of permitted values for $m_{\phi}$ is quite narrow.
Even so, it is remarkable that our preferred range for the 
mass of the boson derived from the dynamics of dSph galaxies
is compatible with those given by other authors 
to ameliorate the problem of overabundance of substructure and is consistent
with the CMB radiation \citep{hu,matos_3,ivan}.

In a recent posting
during the course of submitting this paper, Slepian and Goodman 
\citep{sle11} constrain
the mass of bosonic DM using rotation
curves of galaxies and Bullet Cluster measurements of the scattering 
cross section of self-interacting DM (non bosonic) under the assumption
that these systems are in thermodynamic equilibrium.
If their assumptions are confirmed to be valid, repulsive bosonic DM
will be excluded and, thereby, the only remaining window open
is non-interacting bosons with masses in the range given in equation (\ref{eq:range}).
Nevertheless, the static diffusive equilibrium between Bose-Einstein
condensate and its
non-condensated envelope, as well as finite temperature
effects need to be reconsidered \citep{harko11_2}. 
In addition, other authors \citep{rindler11,ivan2} 
argue that scattering cross sections for bosonic DM are
much smaller than those derived from the condition of thermodynamic 
equilibrium by Slepian and Goodman \cite{sle11}.


\section{Acknowledgments}
The authors wish to thank the referee for useful comments.
V.L.~gratefully acknowledges support from the Alexander von 
Humbold Foundation fellowship and Stu group. 
A.B.~acknowledges CONACyT postdoctoral grant. 
This work was partially supported by CONACyT 
under grant 60526 and by DGAPA-UNAM under grant IN115311. 
This is part of the Instituto Avanzado de Cosmolog\'{\i}a (IAC) 
collaboration.

\appendix
\section{Wake created by a gravitational perturber in an infinite
homogeneous medium}
In order to derive the dynamical friction felt by
a gravitational perturber moving
through a SFDM medium or Bose-Einstein condensate, it is
convenient to use the hydrodynamical representation 
\citep{chavanis1,pitaevskii}.
In the quantum-mechanical equation of motion
(\ref{schroedingerA}), we write the wave function in the form
$\psi(\rr,t)=A(\rr,t) \exp(iS(\rr,t)/\hbar)$ where $A$ and $S$ are real
functions, and make the Madelung transformation
$\rho=m_{\phi}^{2}|\psi|^{2}=m_{\phi}^{2}A^{2}$
and $\vv=\nab S/m_{\phi}$, to find
\begin{equation}
\frac{\partial \rho}{\partial t}+\nab (\rho\cdot \vv)=0,
\label{eq:continuity}
\end{equation}
\begin{equation}
\frac{\partial \vv}{\partial t}+(\vv\cdot\nab)\vv=
-\nab\left(U+U_{\rm ext}+Q\right)
-\frac{1}{\rho}\nab P_{s},
\label{eq:motion}
\end{equation}
where $P_{s}$ is the pressure arising from the short-range interactions,
$U_{\rm ext}$ is the moving and rigid gravitational potential
created by the perturber (a GC in our case)
and
\begin{equation}
Q=-\frac{\hbar^{2}}{2m_{\phi}^{2}\sqrt{\rho}}\nabla^{2}\sqrt{\rho}
\end{equation}
is the quantum potential, which arises from the Heisenberg uncertainty
principle.
The equation of state is polytropic, $P_{s}=K\rho^{\gamma}$,
with index $\gamma=2$ and $K=\lambda/(4m_{\phi}^{4})$ \cite{goodman,chavanis1}.
Thus, we can define the sound speed as
$c_{s}=(2K\rho)^{1/2}$.

The Thomas-Fermi approximation amounts to neglecting the quantum potential
in Eq.~(\ref{eq:motion}). In that case, Equations 
(\ref{eq:continuity})-(\ref{eq:motion})
reduce to the usual Euler equations of a fluid.
Therefore, all we know about dynamical friction in gaseous media
holds valid for a Bose-Einstein condensate.

\begin{figure}
\centerline{\epsfysize=12.5cm \epsfbox{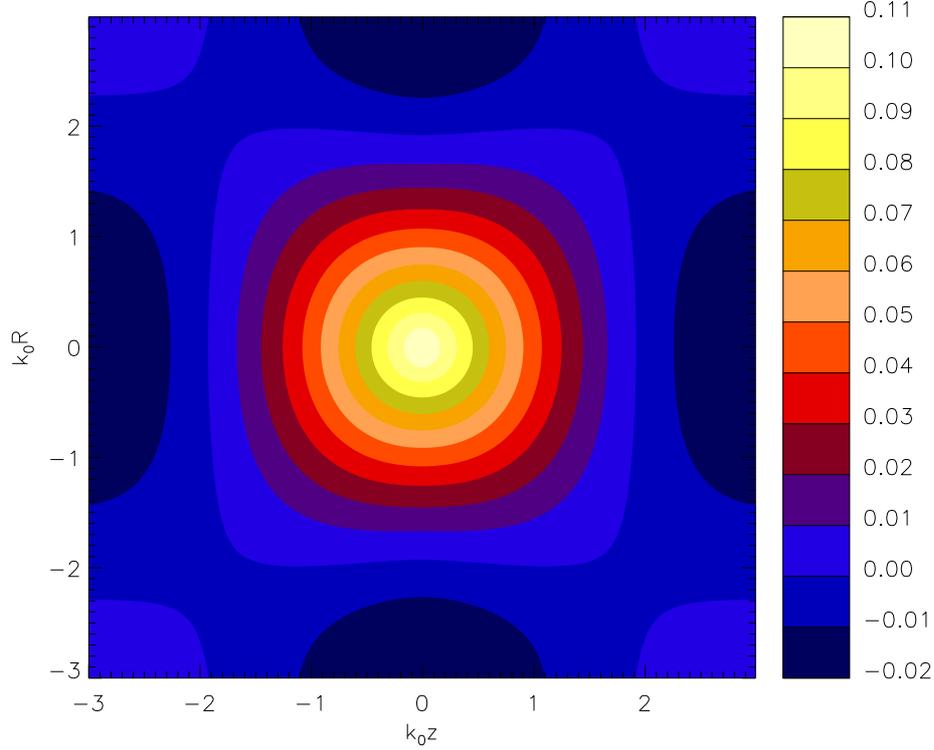}}
\caption{Map of $\xi_{\rm \scriptscriptstyle II}$. Note that
this is a dimensionless quantity that describes the front-back symmetric
part of the perturbed density in units of 
$GM_{p}k_{0}\rho_{0}/V_{0}^{2}$.}
\label{fig:map}
\end{figure}

In the noninteracting case, the particles
interact only via gravity and thus $P_{s}=0$.
We study the completely steady flow created
by a gravitational perturber moving in a straight-line trajectory
at constant-speed $\VV_{0}$ through a homogeneous (infinite) medium. 
The gravitational attraction between the perturber and its own wake
is the dynamical friction force.
Consider a particle at the origin of our coordinate system, surrounded
by a scalar field medium whose velocity far from the particle is $\VV_{0}=-V_{0}
\hat{\zz}$, with $V_{0}>0$.
We will assume that the perturber induces a small perturbation and hence
describe the medium's response in linear theory. Once linearized, the quantum
Euler equations (\ref{eq:continuity})-(\ref{eq:motion}) with $P_{s}=0$
are 
\begin{equation}
\rho_{0}\nab\cdot\vv'+\VV_{0}\cdot\nab\rho'=0,
\label{eq:linear_continuity}
\end{equation}
\begin{equation}
\VV_{0}\cdot\nab\vv'=-\nab U_{\rm ext}+
\frac{\hbar^{2}}{4m_{\phi}^{2}}\nab\left[\nabla^{2}
\frac{\rho'}{\rho_{0}}\right],
\label{eq:linear_motion}
\end{equation}
where $\rho'=\rho-\rho_{0}$ and $\vv'=\vv-\VV_{0}$.
By substituting Eq.~(\ref{eq:linear_continuity}) into the divergence
of Eq.~(\ref{eq:linear_motion}),
we find a single differential equation for the density perturbation,
$\alpha\equiv \rho'/\rho_{0}$,
\begin{equation}
-V_{0}^{2}\frac{\partial^{2} \alpha}{\partial z^{2}}=-\nabla^{2}U_{\rm ext}
+\frac{\hbar^{2}}{4m_{\phi}^{2}}\nabla^{2}\left[\nabla^{2}\alpha\right],
\label{eq:Euler_beam}
\end{equation}
which is a three dimensional version of the Bernoulli-Euler beam equation.
For a point-mass $M_{p}$, we have that $\rho_{\rm ext}=
M_{p}\delta(x)\delta(y)\delta(z)$ and, hence,
$\nabla^{2}U_{\rm ext}=4\pi G M_{p}\delta(x)\delta(y)\delta(z)$.
Taking the Fourier transform of Equation (\ref{eq:Euler_beam}),
we have
\begin{equation}
\frac{\hbar^{2}}{4m_{\phi}^{2}}\left(k^{4}-4k_{0}^{2}k_{z}^{2}\right)
\hat{\alpha}=\sqrt{\frac{2}{\pi}}GM_{p},
\end{equation}
where $k_{0}\equiv m_{\phi}V_{0}/\hbar$.
The density perturbation is recovered by doing the inverse Fourier
transform,
\begin{equation}
\alpha (\rr)=\frac{2GM_{p}m_{\phi}^{2}}{\pi^{2}\hbar^{2}}
\int \frac{\exp(i\kk\cdot\rr)}{k^{4}-4k_{0}^{2}k_{z}^{2}} d^{3}\kk.
\end{equation}
In order to find the poles of the integrand, it is convenient to
write the above equation as
\begin{equation}
\alpha (\rr)=\frac{2GM_{p}m_{\phi}^{2}}{\pi^{2}\hbar^{2}}
\int \frac{\exp(i\kk\cdot\rr)}{(k_{z}^{2}-\chi_{\scriptscriptstyle +}^{2})
(k_{z}^{2}-\chi_{\scriptscriptstyle -}^{2})} d^{3}\kk,
\label{eq:fourier_3D}
\end{equation}
where $\chi_{\scriptscriptstyle \pm}=k_{0}\pm\sqrt{k_{0}^{2}-k_{R}^{2}}$ and
$k_{R}^{2}=k_{x}^{2}+k_{y}^{2}$. The integral (\ref{eq:fourier_3D}) along
$k_{z}$ is evaluated by transforming to the complex plane.
For $k_{R}>k_{0}$, all the poles are outside the contour of integration,
whereas the integrand has poles on the real axis for $k_{R}<k_{0}$.
This reflects the fact that Fourier modes with $k_{R}>k_{0}$ has
a velocity group larger than the perturber's velocity and thus are subsonic,
whereas modes with small wavenumber ($k_{R}<k_{0}$) are supersonic\footnote{See
\cite{chavanis1} for the derivation of the velocity group.}.
It is convenient to subdivide the interval of integration over $k_{R}$ into
two parts: $[0,k_{0}]$ (interval I) and $[k_{0},\infty ]$ (interval II). Thus
$\alpha(\rr)=\alpha_{\rm \scriptscriptstyle I}(\rr)+
\alpha_{\rm \scriptscriptstyle II}(\rr)$, where
$\alpha_{\rm \scriptscriptstyle I}(\rr)$
corresponds to the $k_{R}$-integral over interval I and
$\alpha_{\rm \scriptscriptstyle II}(\rr)$ is the integral over interval II. 
The component $\alpha_{\rm \scriptscriptstyle II}(\rr)$ displays 
front-back symmetry, whereas the contribution
$\alpha_{\rm \scriptscriptstyle I}(\rr)$ is nonzero only behind the body
in order to preserve causality. 
In terms of the dimensionless distance along 
the symmetry axis $\tilde{z}\equiv k_{0}z$, and 
the dimensionless cylindrical radius $\tilde{R}\equiv k_{0}R$,
the density perturbation can be written as
\begin{equation}
\alpha(\tilde{z},\tilde{R})=\frac{GM_{p}k_{0}}{V_{0}^{2}}
\left[\xi_{\rm \scriptscriptstyle I}(\tilde{z},\tilde{R})+
\xi_{\rm \scriptscriptstyle II}(\tilde{z},\tilde{R})\right],
\end{equation}
where $\xi_{\rm \scriptscriptstyle I}(\tilde{z},\tilde{R})$
and $\xi_{\rm \scriptscriptstyle II}(\tilde{z},\tilde{R})$ are dimensionless
quantities, and
the subscripts I and II refer to the asymmetric and the symmetric parts,
respectively. We see that the density enhancement at $(\tilde{z},\tilde{R})$
decreases with increasing $V_{0}$ and increases with $k_{0}$.

For illustration, the symmetric part of the
density perturbation, $\xi_{\rm \scriptscriptstyle II}$,
is shown in Figure \ref{fig:map}.  This map represents the
symmetric contribution of the steady-state density perturbation
created by a gravitational perturber moving in an infinite homogeneous 
medium.
We see that the density isocontours are extremely
spherical (for a comparison with the isocontours in a gaseous medium
see \cite {ostrikereve}). 
The time-dependent evolution of the wake generated by,
and the gravitational drag force on, a body traveling on 
a circular orbit will be presented in a forthcoming paper.



\end{document}